\documentclass[sigconf,table,screen]{acmart}

\AtBeginDocument{%
  \providecommand\BibTeX{{%
    \normalfont B\kern-0.5em{\scshape i\kern-0.25em b}\kern-0.8em\TeX}}}

\setcopyright{none}
\renewcommand\footnotetextcopyrightpermission[1]{}
 \acmConference[ESEC/FSE '22]{Proceedings of the 30th ACM Joint European Software Engineering Conference and Symposium on the Foundations of Software Engineering}{November 14--18, 2022}{Singapore, Singapore}

\usepackage{balance}
\usepackage{flushend}

\usepackage{graphicx}
\usepackage{subcaption}
\usepackage{enumitem}
\usepackage[shortcuts]{extdash}
\usepackage{xcolor}
\usepackage{arydshln}
\usepackage{enumitem}
\usepackage{xspace}
\usepackage{todonotes}
\usepackage{indentfirst}
\usepackage{ifthen}
\usepackage{hyperref}
\hypersetup{
	colorlinks   = true,
	citecolor    = blue,
	linkcolor    = blue,
	urlcolor=blue
}
\usepackage{cleveref}
\usepackage{graphics}
\usepackage{multirow}
\usepackage{makecell}
\usepackage{colortbl}
\usepackage{tcolorbox}
\usepackage[shortcuts]{extdash}

\usepackage{adjustbox}

\begin{document}

\title[Software Security during Modern Code Review: The Developer's Perspective]{Software Security during Modern Code Review: \\ The Developer's Perspective}

\author{Larissa Braz}
\affiliation{%
  \institution{University of Zurich}
  \country{Switzerland}
  }
\email{larissa@ifi.uzh.ch}

\author{Alberto Bacchelli}
\affiliation{%
  \institution{University of Zurich}
  \country{Switzerland}
  }
\email{bacchelli@ifi.uzh.ch}


\newcommand{\etal}{\textit{et al.}\xspace}
\newcommand{\eg}{\textit{e.g.,}\xspace}
\newcommand{\ie}{\textit{i.e.,}\xspace}

\newcommand{\surveyStarted}{\bluetext{329}\xspace}
\newcommand{\surveyNotFinished}{\bluetext{121}\xspace}

\newcommand{\totalInterviews}{10\xspace}
\newcommand{\totalInterviewsWritten}{ten\xspace}

\newcommand{\sda}{\emph{SDA}\xspace}
\newcommand{\dcs}{\emph{DCS}\xspace}

\newcommand{\interviewee}[1]{\textsc{I#1}\xspace}
\newcommand{\surveyR}[1]{\textsc{R#1}\xspace}

\newcommand{\csunits}{\textcolor{red}{XXX}\xspace}

\newcommand{\rqOne}{What is the current developer’s perspective on ensuring software security during code review?\xspace}
\newcommand{\rqTwo}{To what extent do companies/projects support security assessment during code review?\xspace}
\newcommand{\rqThree}{What are the main challenges experienced when
	ensuring security during modern code reviews?\xspace}

\definecolor{gray50}{gray}{.5}
\definecolor{gray40}{gray}{.6}
\definecolor{gray30}{gray}{.7}
\definecolor{gray20}{gray}{.8}
\definecolor{gray10}{gray}{.9}
\definecolor{gray05}{gray}{.95}

\definecolor{purple}{HTML}{DADAEB}
\definecolor{blue1}{HTML}{e1effc}
\definecolor{babypink}{HTML}{ffedf8}
\definecolor{palevioletred}{HTML}{db7093}

\newlength\Linewidth
\def\findlength{\setlength\Linewidth\linewidth
	\addtolength\Linewidth{-4\fboxrule}
	\addtolength\Linewidth{-3\fboxsep}
}

\newenvironment{rqbox}{\par\begingroup
	\setlength{\fboxsep}{5pt}\findlength
	\setbox0=\vbox\bgroup\noindent
	\hsize=0.95\linewidth
	\begin{minipage}{0.95\linewidth}\normalsize}
	{\end{minipage}\egroup
	\textcolor{gray20}{\fboxsep1.5pt\fbox
		{\fboxsep5pt\colorbox{purple}{\normalcolor\box0}}}
	\endgroup\par\noindent
	\normalcolor\ignorespacesafterend}
\let\Examplebox\examplebox
\let\endExamplebox\endexamplebox

\definecolor{light-purple}{HTML}{dadaeb}
\newcommand{\rb}[1]{
	\begin{tcolorbox}[colback=purple,
		colframe=black,
		width=\columnwidth,
		arc=3mm, auto outer arc,
		boxrule=0.5pt,
		]
		#1
	\end{tcolorbox}
	\vspace{0.1cm}
}

\newcounter{Observation}
\stepcounter{Observation}

\newcommand{\roundedbox}[1]{
	\rb{
		\noindent
		\textit{\textbf{Observation \theObservation}. #1}
	}
	\stepcounter{Observation}
}


\newboolean{showcomments}
\setboolean{showcomments}{true}
	
\ifthenelse{\boolean{showcomments}}
{
	\newcommand{\nb}[3]{
		{\colorbox{#2}{\bfseries\sffamily\scriptsize\textcolor{white}{#1}}}
		{\textcolor{#2}{\textsf\small$\blacktriangleright$\textit{#3}$\blacktriangleleft$}}}
	 \newcommand{\version}{\emph{\scriptsize$-$Id$-$}}
}{
	\newcommand{\nb}[3]{}
} 

\newcommand{\hide}[1]{}

\newcommand{\ab}[1]{\nb{Alberto}{palevioletred}{#1}}
\newcommand{\lari}[1]{\textcolor{magenta}{#1}}

\newcommand{\bluetext}[1]{\textcolor{black}{#1}}

\newcommand{\numComplete}{\bluetext{208}\xspace}
\newcommand{\numParticipantsSurvey}{182\xspace}
\newcommand{\numParticipantsNoCR}{24\xspace}
\newcommand{\numEmployed}{178\xspace}
\newcommand{\numStudents}{1\xspace}
\newcommand{\numIndependentContractor}{12\xspace}
\newcommand{\numUnemployed}{13\xspace}
\newcommand{\devs}{142\xspace}
\newcommand{\perDevs}{78\%\xspace}
\newcommand{\numDevsEmbbeded}{5\xspace}
\newcommand{\numDevsFront}{2\xspace}
\newcommand{\numDevsBack}{45\xspace}
\newcommand{\numDevsFull}{78\xspace}
\newcommand{\numDevsMobile}{2\xspace}
\newcommand{\numDevsOther}{10\xspace}
\newcommand{\totalExperienceOverThree}{153\xspace}
\newcommand{\perExperienceOverThree}{84\%\xspace}
\newcommand{\totalOftenCR}{113\xspace}
\newcommand{\perOftenCR}{62\%\xspace}
\newcommand{\totalOftenDesign}{123\xspace}
\newcommand{\perOftenDesign}{68\%\xspace}
\newcommand{\totalOftenProgramming}{166\xspace}
\newcommand{\perOftenProgramming}{91\%\xspace}
\newcommand{\totalCRvalue}{167\xspace}
\newcommand{\totalCRtold}{15\xspace}
\newcommand{\perCRvalue}{92\%\xspace}
\newcommand{\perCRtold}{8\%\xspace}
\newcommand{\totalIssuesSecurity}{9\xspace}
\newcommand{\totalIssuesSecurityWritten}{nine\xspace}
\newcommand{\totalIssuesBugs}{37\xspace}
\newcommand{\totalIssuesPerformance}{23\xspace}
\newcommand{\perPerformance}{13\%\xspace}
\newcommand{\totalFamiliarSecurity}{145\xspace}
\newcommand{\totalNotFamiliarSecurity}{37\xspace}
\newcommand{\totalReported}{118\xspace}
\newcommand{\totalAlwaysConsiderAgreeOrStrongly}{111\xspace}
\newcommand{\totalSensitiveAgreeOrStrongly}{105\xspace}
\newcommand{\perSensitiveAgreeOrStrongly}{58\%\xspace}
\newcommand{\totalCanDecideAgreeOrStrongly}{148\xspace}
\newcommand{\perCanDecideAgreeOrStrongly}{81\%\xspace}
\newcommand{\totalAlwaysAgreeOrStronglyMentionedSecurity}{nine\xspace}
\newcommand{\totalAffectedStrongOrAgree}{147\xspace}
\newcommand{\perAffectedStrongOrAgree}{81\%\xspace}
\newcommand{\totalSecReviewStrongOrAgree}{21\xspace}
\newcommand{\perSecReviewtrongOrAgree}{12\%\xspace}
\newcommand{\totalLookStrongOrAgree}{131\xspace}
\newcommand{\perLookStrongOrAgree}{81\%\xspace}
\newcommand{\totalDevRespStrongOrAgree}{126\xspace}
\newcommand{\perDevRespOrAgree}{69\%\xspace}
\newcommand{\totalChallCRStrongOrAgree}{92\xspace}
\newcommand{\perChallCROrAgree}{51\%\xspace}
\newcommand{\totalChallAllStrongOrAgree}{67\xspace}
\newcommand{\perChallAllRespOrAgree}{37\%\xspace}
\newcommand{\totalHasSecurityTeam}{85\xspace}
\newcommand{\perHasSecurityTeam}{47\%\xspace}
\newcommand{\totalACK}{43\xspace}
\newcommand{\perACK}{24\%\xspace}
\newcommand{\totalNotACK}{122\xspace}
\newcommand{\perNotACK}{67\%\xspace}
\newcommand{\totalDoMore}{149\xspace}
\newcommand{\perDoMore}{82\%\xspace}
\newcommand{\totalMainConcern}{81\xspace}
\newcommand{\perMainConcern}{45\%\xspace}
\newcommand{\totalProvidesSecurity}{70\xspace}
\newcommand{\perProvidesSecurity}{38\%\xspace}
\newcommand{\totalAllowsTraining}{95\xspace}
\newcommand{\perAllowsTraining}{52\%\xspace}
\newcommand{\totalChallLackKnowledge}{44\xspace}
\newcommand{\totalChallCRlimitations}{five\xspace}
\newcommand{\totalChallAwareness}{eight\xspace}
\newcommand{\totalChallCodebase}{nine\xspace}
\newcommand{\totalChallThird}{eight\xspace}
\newcommand{\totalChallSubtle}{nine\xspace}
\newcommand{\totalChallTime}{six\xspace}
\newcommand{\totalChallInterest}{one\xspace}
\newcommand{\totalChallResources}{two\xspace}
\newcommand{\totalChallMitTraining}{thirty-two\xspace}
\newcommand{\totalChallMitTool}{15\xspace}

\begin{abstract}
To avoid software vulnerabilities, organizations are shifting security to earlier stages of the software development, such as at code review time.
In this paper, we aim to understand the developers’ perspective on assessing software security during code review, the challenges they encounter, and the support that companies and projects provide. 
To this end, we conduct a two-step investigation: we interview \totalInterviews professional developers and survey \numParticipantsSurvey practitioners about software security assessment during code review. 
The outcome is an overview of how developers perceive software security during code review and a set of identified challenges.
Our study revealed that most developers do not immediately report to focus on security issues during code review. 
Only after being asked about software security, developers state to always consider it during review and acknowledge its importance. Most companies do not provide security training, yet expect developers to still ensure security during reviews. 
Accordingly, developers report the lack of training and security knowledge as the main challenges they face when checking for security issues. In addition, they have challenges with third-party libraries and to identify interactions between parts of code that could have security implications. Moreover, security may be disregarded  during reviews due to developers' assumptions about the security dynamic of the application they develop.
%


\noindent\textbf{Data and materials:} \url{https://doi.org/10.5281/zenodo.6875435}

\end{abstract}

\ccsdesc[500]{Security and privacy~Software security engineering}

\keywords{code review, security, software vulnerabilities}

\maketitle

\newpage
\section{Introduction}

A software vulnerability is a security flaw, glitch, or weakness found in software code that could be exploited by an attacker~\cite{dempsey:2017} to cause harm to the stakeholders of a software system~\cite{owasp}. To avoid vulnerabilities in software systems, organizations are shifting security ``left,'' that is, to earlier stages of software development, such as during code review~\cite{gitlab-survey}. 
Code review is a widely agreed-on practice~\cite{Boehm:2001} recognized as a valuable tool for reducing software defects and improving the quality of software projects~\cite{Ackerman:1984, Ackerman:1989,Bacchelli:2013}. 
Previous studies show that code review is also an important practice for detecting and fixing security bugs earlier~\cite{Thompson:2017,McGraw:2004} and has positive effects on Secure Development of Applications (\sda)~\cite{Meneely:2012,Shin:2011,Meneely:2010}. 
An application that has not been reviewed for security gaps is likely to have problems virtually at 100\%~\cite{owaspChecklist}.

Previous studies~\cite{woon:2007,Xie:2011,Xiao:2014,Poller:2017} have reported and investigated the limited adoption of \sda and investigated the developer's perspective on security. However, as software security seems to increasingly gain more space among developers’ responsibilities, further studies are needed to investigate this left-shift. In particular, in this paper, we focus on the current state of security during code review according to practitioners.
\bluetext{Our definition of security is rooted in the definition of software vulnerabilities. In fact, ensuring security during code review means finding and proposing fixes for security vulnerabilities in the code under review to avoid turning the application insecure~\cite{owaspChecklist}. %
Moreover, }findings on this topic can provide insight for practitioners, organizations, and researchers. Developers and other software project stakeholders can use our insights to improve their code review practices to better ensure security during code review. Organizations can use evidence about developers' challenges during code reviews to update and improve their approaches and methodologies toward achieving and maintaining secure applications. Finally, researchers can focus their attention on developers’ challenges to facilitate secure code review.

We set up our investigation as an explorative study. Thus, instead of starting with preset hypotheses on how software security is or should be addressed during code reviews, we investigate: (1) the developer’s perspective on assessing software security during code review; (2) the expectations and support that companies and projects have on this process; and (3) what problems software engineers face when evaluating software security during reviews.

We structure our investigation into two steps. In the first step, we interview ten professional developers by means of 30 minutes semi-structured interviews. In the second step, we design and disseminate an online survey to collect developers' perceptions, practices, and challenges concerning software security inspection during modern code review. 
The survey received a total of \numParticipantsSurvey valid answers. Among the survey respondents, \perDevs (\devs) report being software developers currently, \perExperienceOverThree (\totalExperienceOverThree) have three years or more of professional development experience, and \perOftenCR (\totalOftenCR) conduct code reviews at least several times per week.

Our study reveals that most participants (\totalSensitiveAgreeOrStrongly) develop security-sensitive software systems; yet, when asked which issues they focus on during code review, only \totalIssuesSecurity survey respondents explicitly mention security. Only after we mention security, \totalAlwaysConsiderAgreeOrStrongly respondents state to \emph{always} consider it during code review. 
This stark mismatch may indicate that security is not a priority during review and may be assessed less frequently than reported.
Moreover, developers may disregard security aspects during reviews due to their assumptions about the security dynamic of the application they develop.

Concerning companies, most respondents (\totalDevRespStrongOrAgree) work for companies or projects that expect developers to ensure security, including through code review. However, the vast majority of the respondents (\totalDoMore) think that companies should do more to support secure practices. For example, \totalNotACK respondents are not acknowledged for performing secure code reviews. Moreover, two-thirds of the respondents state that companies do not provide security training, rather just allow developers to acquire security skills by themselves during working hours (reported by \totalAllowsTraining respondents).

Indeed, when it comes to challenges of ensuring security in code review, respondents mostly report lack of training and knowledge as the main issue they face (mentioned by \totalChallLackKnowledge respondents). Second, the use of third-party libraries is problematic because they are not always present in the code developers maintain; thus, checking them actively during code review is not possible (mentioned by \totalChallThird respondents). Finally, some respondents explained that code review is not the ideal phase of the development process to detect vulnerabilities. For instance, some vulnerabilities require the execution of the code to appear. Moreover, developers find it hard to identify interactions among parts of code that could have security implications at low-level inspections.

Based on the findings of our study, we propose a series of recommendations and areas for future research.

\section{Background and Related Work}

In this section, we review the Software Engineering (SE) research literature on topics investigating software security and the developer-related factors that influence vulnerability detection.

\smallskip
\noindent\textbf{Security in the Development Lifecycle. } 
In 2018, the McAfee institution~\cite{mcafee-report} reported that, daily, 80 billion malicious scans look for vulnerable targets, and 780 thousand records are lost to hacking. However, security experts still have to motivate and convince developers of the importance of finding vulnerabilities~\cite{Thomas:2018}.
Later, GitLab performed a survey with over 3,650 respondents from 21 countries~\cite{gitlab-survey}. In their survey, security experts reported that it is a software developer's job to develop secure code, but only less than half of the developers can actually detect vulnerabilities.

Security often fails because users either misunderstand the security implications of their actions or turn off security features to workaround usability problems~\cite{Balfanz:2004}.
In fact, security usabi\-lity issues also affect developers. For instance, storing user login data and authenticating users is prone to security issues due to the high complexity of the technologies and concepts involved in the process~\cite{Green:2016}. Usability issues existing in such security APIs force non-security expert programmers to misuse these APIs and introduce security vulnerabilities
to applications they develop~\cite{Wijayarathna:2018}.
Developer-Centred Security (\dcs) studies have addressed some of the developers' needs and attempted to apply existing Human-Computer Interaction methodologies and to adopt well\-/established usable security measures to software development~\cite{Wurster:2008,Green:2016,Pieczul:2017}.
However, \citet{Tahaei:2019} report a lack of research on several aspects of \dcs, including how to make security a business value and security often being ignored because it is a secondary requirement. 

In 2004, Microsoft released a Trustworthy Computing Security Development Lifecycle initiative~\cite{microsofttrust}, a process adopted by the company to develop software that needs to withstand malicious attacks. The process included the addition of a series of security-focused activities to each phase of Microsoft's software development process. 
\citet{lipner:2004} investigated the effectiveness of Microsoft's initiative and found that the practice of Secure Development of Applications (\sda) provides security benefits.

In 2007, \citet{woon:2007} conducted a field survey of 184 information system professionals to investigate the factors that may influence the intention of the participants to practice Secure Development of Applications (\sda), \ie incorporate security as part of the application development lifecycle. Their results show evidence that participants' intention was determined primarily by attitude, followed by product usefulness and subjective norm. Moreover, self-efficacy and facilitating conditions did not appear to impact the information system professionals' intention to practice \sda. In addition, participants reported that companies did not facilitate \sda practice other than allowing the respondents to attend seminars on the topic. Professionals could decide whether to practice it, but the organizations did not give them rewards or recognition for their extra efforts.
Moreover, \citet{woon:2007} reported limited adoption of \sda and a lack of studies exploring the phenomenon.
Later, \citet{Xie:2011} identified an ``it is not my responsibility'' attitude from the developers towards \sda.
Several studies~\cite{Assal:2018,Xiao:2014,Poller:2017} found that developers prioritize more-visible functional requirements or even easy\-/to\-/measure activities, \eg closing bug tracking tickets, over security. 
On the contrary, \citet{Christakis:2016} reported that developers care more about security than other reliability issues. 

\citet{Smith:2018} advocate that static analysis tools detect vulnerabilities and help developers resolve them. Previous studies have proposed and improved tooling support according to developers’ needs~\cite{Ayewah:2008,Ayewah:2008-2,Smith:2015}, but tools are still generally poorly adopted by developers~\cite{Tahaei:2019} as they are confusing for developers to use~\cite{Smith:2018}.

The aforementioned studies~\cite{Xie:2011,woon:2007, Christakis:2016, Assal:2018, Xiao:2014, Poller:2017} provide insights on \sda and how developers perceive security. However, they might not represent the current development scenario. For instance, \citet{woon:2007} study was conducted 15 years ago, and \cite{Xie:2011} was performed 11 years ago. Nowadays, as software security seems to be gaining more space among developers’ responsibilities, in this paper, we investigate whether their findings still hold. To this aim, we investigate the current developer's perspective on this phenomenon and whether they assess software security during code reviews. We created our initial interviews guideline according to the findings of these studies. 

\smallskip
\noindent\textbf{Code Review and Security. } 
Code review is a way to manually inspect source code by developers other than the author~\cite{Cohe2010a}. In its contemporary practice, code review is asynchronous, tool- and change-based~\cite{baum2016factors}, and widely used across companies~\cite{Bacchelli:2013,sadowski2018modern} as well as community-driven projects~\cite{rigby2013convergent,rigby2014peer}. \citet{Bacchelli:2013} surveyed developers and managers' expectations regarding code review at Microsoft. 
Among the observations they collected during interviews, a senior developer stated: ``I’ve seen quite a few code reviews where someone commented on formatting while missing the fact that there were security issues.''

\citet{edmundson:2013} stated that manual code review could be expensive and impractical due to the need for several reviewers to inspect a piece of code to find a vulnerability. 
On the contrary, \citet{Weir:2016} interviewed twelve industry experts to investigate how to improve the security skills of mobile app developers. Some of the techniques recommended included code reviews.
In this vein, the OWASP institution~\cite{owaspChecklist} defined secure code review as probably the single-most effective technique for identifying security bugs early in the system development lifecycle.

\citet{braz:2021} investigated to what extent software developers can detect vulnerabilities during code reviews. They found that several developers often miss a popular and easy-to-detect vulnerability when reviewing code; yet, when explicitly informed about the presence of a vulnerability in the change, a significant portion of the additional developers could identify it.
Later, \citet{braz:2022} investigated whether and to what extent instructing developers to focus on security issues and providing security checklists during code reviews can support the detection of software vulnerabilities. They found that developers' mental attitude plays a role in detecting software vulnerabilities during code reviews. The effect of security instructions provided evidence that vulnerability detection could be triggered with proper security considerations, such as security standards for code reviews. 

These studies~\cite{braz:2021,braz:2022} have investigated the developer's ability to detect vulnerabilities during a practical code review task. They also provided the first evidence on how developers could perceive security assessment during code review. In this work, we continue on this research path, focusing specifically on developers' perceptions as a way to generate knowledge to inform current practices and research.

\section{Methodology}\label{sec:method}

Overall, our research aims to understand the developers’ perspective on assessing software security during code review and the support companies and projects provide to this process. We base our study on semi-structured interviews and a survey we devised and disseminated to collect different types of self-reported data.

\subsection{Research Questions}
\label{method:rqs}

Our investigation is structured around three research questions. We incorporated them in the interview guidelines and the survey.

As software security seems to be gaining more space among developers’ responsibilities, we first investigate the developer's perspective on this phenomenon and whether the developers in fact assess software security during code reviews: 

\smallskip
\begin{center}
	\begin{rqbox}
		\begin{description}	    
			\item[]\textbf{RQ$_1$.} \emph{\rqOne}
		\end{description} 
	\end{rqbox}
\end{center} 

Organizations are shifting security to earlier stages of software development, increasing developers' responsibility around security. %
We investigate the support that companies and projects provide to the software security assessment during code reviews:

\smallskip
\begin{center}
	\begin{rqbox}
		\begin{description}	    
			\item[]\textbf{RQ$_2$.} \emph{\rqTwo}
		\end{description} 
	\end{rqbox}
\end{center} 

Finally, we focus on the challenges developers face when ensuring software security during code reviews. 
Such an understanding helps us obtain a catalog of reasons why this practice is not adopted and allow us to propose recommendations to mitigate them.

\smallskip
\begin{center}
	\begin{rqbox}
		\begin{description}	    
			\item[]\textbf{RQ$_3$.} \emph{\rqThree}
		\end{description} 
	\end{rqbox}
\end{center}

\subsection{Semi-structured Interview Design}
\label{method:interviews}

We performed the first part of our study as a set of one-to-one semi-structured interviews~\cite{lindlof:2002} with professional software developers of different backgrounds. Semi-structured interviews use of an interview guide that contains general groupings of topics and questions rather than a pre-determined set and order of questions~\cite{lindlof:2002,Bacchelli:2013}. They are often used in an exploratory context to
``find out what is happening [and] to seek new insights''~\cite{weiss:1995}. The guideline was iteratively refined after each interview, particularly when developers started providing answers very similar to the earlier ones, thus reaching a saturation effect.
The main path of the final interview guideline has 28 questions. \Cref{tab:interview-questions} shows two of the questions we asked participants during the interviews. 
Following, we describe the guideline's main points. 


\begin{table}[b]
	\centering 
	\caption{Sample of the interviews questions.} 
	\label{tab:interview-questions}
	\begin{tabular}{ll}
		\hline
		Q15 & To what extent do you consider security aspects when \\
		& reviewing code?  \\ 
		Q17 & Do you think ensuring security during code reviews is \\
		& challenging? If yes, why do you think it is challenging? \\ 
		\hline
	\end{tabular}
\end{table}

\smallskip
\noindent\textbf{(1) Introduction and background questions. } In the introduction, we spend a few minutes explaining who we are and our research --- \textit{without mentioning security}. We review the main points of the consent form they signed prior to the interviews and answer any questions they might have. For instance, we remind them they were allowed to stop the interview at any moment and also skip any question they did not feel like answering. We also explain that there were no right or wrong answers. Following, we start recording the interview and  ask the participants questions about their background, such as their current role and how many years of experience in programming they have. The discussion of participant background serves as an icebreaker and also provides some context for later in the interview.

\smallskip
\noindent\textbf{(2) Code review questions. } In this part of the interview, we ask general questions about code reviews. First, we ask participants if they perform code reviews as part of their current job or whether they have previous experience with it. Although we explicitly state the interview is about code reviews when inviting participants, it could happen that a participant answer they have no experience with this activity whatsoever. In this case, we move the interview to the wrap-up bulk. Participants that confirm they have conducted code review sessions continue to answer questions about their experiences. For instance, we ask the participant why they perform code reviews and what type of issues they look for when reviewing. At this moment, we observe whether participants mention security or vulnerabilities by themselves, \ie whether security is spontaneously mentioned without being primed.  

\smallskip
\noindent\textbf{(3) Security questions. } We ask participants if they are familiar with the term ``software vulnerabilities'' and their definition of it. For participants unfamiliar with the term, we read OWASP's~\cite{owasp} definition and ask if they agree with it. \bluetext{The subsequent questions and discussion regarding security are in this context of vulnerability detection.} We ask participants whether they consider security when reviewing code. To those who confirm they do it, we ask security-specific questions about their practices and experiences. To participants that do not consider security during reviews, we ask questions about security in other activities of the software development process. We then ask participants about the challenges to ensure security in their reviews and other development activities.

\smallskip
\noindent\textbf{(4) Wrap-up. } We conclude the interviews with some basic questions, for instance: ``Is there anything you want to tell us about code reviews and security that we did not cover before?'' and ``Is there anything you want to ask us?''. Finally, we thank them.

\smallskip
We conducted the interviews through a video-conferencing application. With consent, we recorded the interviews, assuring the participants of anonymity. The audio of each interview was then transcribed and broken up into smaller coherent units for subsequent analysis. The transcripts of the interviews are available~\cite{replication-package}.

\subsection{Survey Design}
\label{sec:methog:survey}

In the second part of our study, we conducted a survey with developers who have experience with code reviews. Our goal was to validate and expand on the findings collected through the semi-structured interviews, and further answer our research questions (see \cref{method:rqs}). We built and ran the survey on Qualtrics~\cite{Qualtrics}. 

The survey had 22 questions \bluetext{compulsory questions, organized into four open-ended questions, three multiple-choice, 15 single-choice, and three 5-point Likert scale grids. These  grids asked participants to rate 8, 7, and 4 items on a 5-point Likert scale (ranging from `strongly disagree' to `strongly agree').}
The multiple-choice and rate questions had randomized answer options and statements. \Cref{tab:survey-questions} shows an example of a closed and an open-ended question. In the following, we detail the survey's design and how the participant flows through each block of questions. Each block corresponds to at least one different page, and returning to previous pages is not allowed. The complete survey is available in the accompanying material~\cite{replication-package}.


\begin{table}[b]
	\centering 
	\caption{Sample of the survey questions.} 
	\label{tab:survey-questions}
	\begin{tabular}{ll}
		\hline
		Q6       & What types of issues do you focus on during your code\\
		            & review sessions? \\ 
		Q12.4 & It is part of the developer's job to ensure the security \\
		             & of the application \emph{(strongly disagree, disagree, neither} \\
		             & \emph{agree nor disagree, agree, strongly agree, not applicable}). \\
		\hline
	\end{tabular}
\end{table}

\smallskip
\noindent\textbf{(1) Welcome page. } On the first page of the survey, we provide participants with information about the study. We do not inform the participants about the study's final focus on software vulnerabilities to avoid they subconsciously changing their answers to fit it (\ie demand characteristics~\cite{Nichols:2008}). We inform participants about the data handling policy, ask for their consent to use their data, and inform them that they are allowed to drop out at any time.

\smallskip
\noindent\textbf{(2) Qualification questions. } We ask participants two random technical questions to screen non-programmers out of our study. We selected three technical questions from \citet{Danilova:2021}'s list as they offer only minimal overhead for the participants with actual programming skills. 

\smallskip
\noindent\textbf{(3) Code review experience and practices. } In this block, we ask questions to gather information about their code reviews' experiences and practices, including the reasons why they perform the reviews and the type of issues they have reported in their last code review sessions. At this point, \emph{we do not ask security-specific questions}. This way, we can identify participants who spontaneously report to include security as part of their code reviews. 
 
\smallskip
\noindent\textbf{(3) Security knowledge and practices during code reviews. } In this step, we ask participants whether they are familiar with the term \textit{software vulnerabilities} and provide them with its definition~\cite{owasp}. \bluetext{The subsequent questions regarding security are in this context of vulnerability detection.} We ask questions to gather information about factors that may affect how the participants address security during code reviews, such as their security knowledge, practices, and company and team culture. Most of the questions are closed in a Likert scale format. Depending on a participant’s answers, some questions were filtered and not presented to avoid asking unnecessary questions. \bluetext{Different from the interview, in the survey we did not explicitly ask participants about the security process followed within their companies. We excluded this question to avoid overloading survey respondents and to increase response quality~\cite{galesic2009effects}.}

\smallskip
\noindent\textbf{(4) Demographics. } We ask participants questions to collect demographic information and confounding factors, such as the highest obtained education and years of professional experience (all questions are available in the replication package~\cite{replication-package}). This information is mandatory to fill in as collecting it helps us identify which portion of the developer population is represented by our respondents~\cite{Falessi:2018}. 

\smallskip
\noindent\textbf{(5) Feedback and closing. } We ask for the participant's feedback on the survey. We also ask if they would like to share their anonymous data in a public research dataset and receive the study results.

\subsection{Interviews and Open Answers Analysis}

We performed two open card sortings~\cite{spencer2009card} to extract emerging themes from the interviews and open answers of the survey. This allowed organizing the codes into hierarchies to deduce a higher level of data abstraction. We followed the same process in both card sortings:  the first authors and an external researcher created self-contained units, then sorted them into themes. To ensure the themes' integrity, the authors interactively sorted the units several times. After review by the last author and discussions, we reached the final themes. Through the discussions, we evaluated controversial answers, reduced potential bias caused by wrong interpretations of a participant's answer, and strengthened the confidence in the card sort output (available in our replication package~\cite{replication-package}).

\subsection{Recruiting Participants}

To recruit participants for the interviews, we contact our professional networks as well as ask for participation on practitioners' web forums. We disseminated the online survey out through practitioners' web forums, IRC communication channels, direct authors' contacts from their professional networks, as well as their social media accounts (\eg Twitter, Facebook). 
We did not reveal the security angle of the study; instead, we explained that the study was about expectations and practices in code review. We also introduced a donation-based incentive of 20 USD and 2 USD to a charity per participant of the interviews and the survey, respectively.

\section{Results}

In this section, we report the results of our investigation.

\subsection{Participants}

In total, we interviewed \totalInterviewsWritten software developers with \bluetext{up to 13} years of industry experience.
Among the interviewees, three also have a technical lead position within their team. 
\bluetext{All interviewees work in different companies of distinct dimensions: three startups, three mid-sized software engineer companies, and four large software engineer companies. One interviewee (\interviewee{1}) is also active in the open source community. Two interviewees are located in Brazil, one in Canada, one in England, one in Portugal, one in the Netherlands, and four in Switzerland.} 
All interviewees perform code reviews at least three times a week. \Cref{tab:interviewees} summarizes the background of the interviewees.

A total of \surveyStarted people started our online survey through the provided link. Of these, \surveyNotFinished did not complete it; thus, we removed their entries from the results dataset. A total of \numComplete people completed all survey blocks. 
We removed \numParticipantsNoCR respondents who reported not performing code reviews and manually inspected five respondents who incorrectly answered at least one qualification question, removing other two respondents. After applying these exclusion criteria, data from \numParticipantsSurvey respondents could be used for the analyses.


\begin{table}[t]
	\centering 
	\caption{Profiles of the interviewed professional developers.} 
	\label{tab:interviewees}
	\begin{tabular}{rllrl}
		\hline
		     &              & \textbf{Company/} & \textbf{Experience} & \textbf{Security}\\
		\textbf{ID} & \textbf{Domain} & \textbf{Project} & \textbf{(in years)} & \textbf{Expert}\\
		\hline
		\emph{I1} & Indust./OSS & Mid SW Comp. & 9  & \multicolumn{1}{c}{\checkmark}\\
		\emph{I2} & Industry & Startup & 1 & \\
		\emph{I3} & Industry  & Large SW Comp. & 1 & \\
		\emph{I4} & Industry  & Large SW Comp. & 11 & \\
		\emph{I5} & Industry  & Startup & 1 & \\ 
		\emph{I6} & Industry  & Large SW Comp. & 13 & \\
		\emph{I7} & Industry  & Mid SW Comp. & 6 & \\
		\emph{I8} & Industry  & Mid SW Comp. & 7 & \\
		\emph{I9} & Industry  & Startup & 2 & \\
		\emph{I10} & Industry & Large SW Comp. & 8 & \\
		\hline
	\end{tabular}
\end{table}

\begin{figure}[b]
	\centering
	\includegraphics[width=1\columnwidth,]{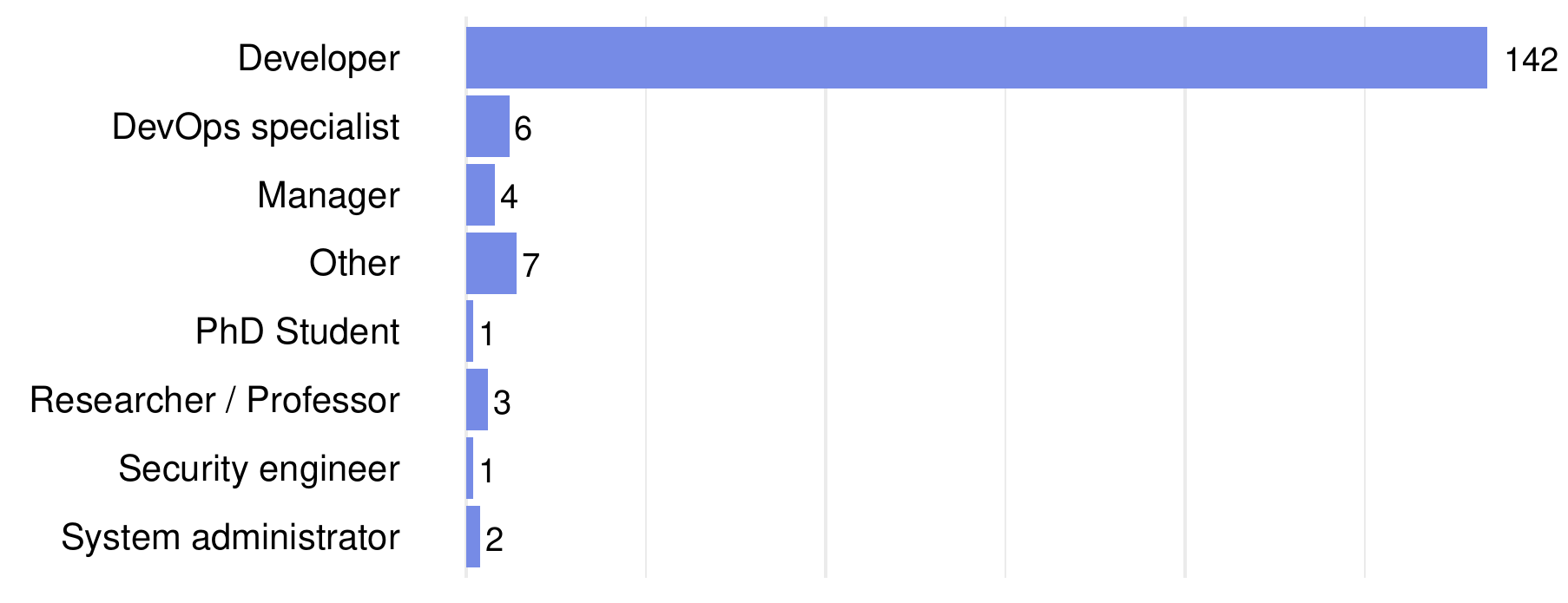}
	\caption{Job distribution among employed respondents.}
	\label{fig:jobs}
\end{figure}

\begin{figure}[b]
	\includegraphics[width=1\columnwidth]{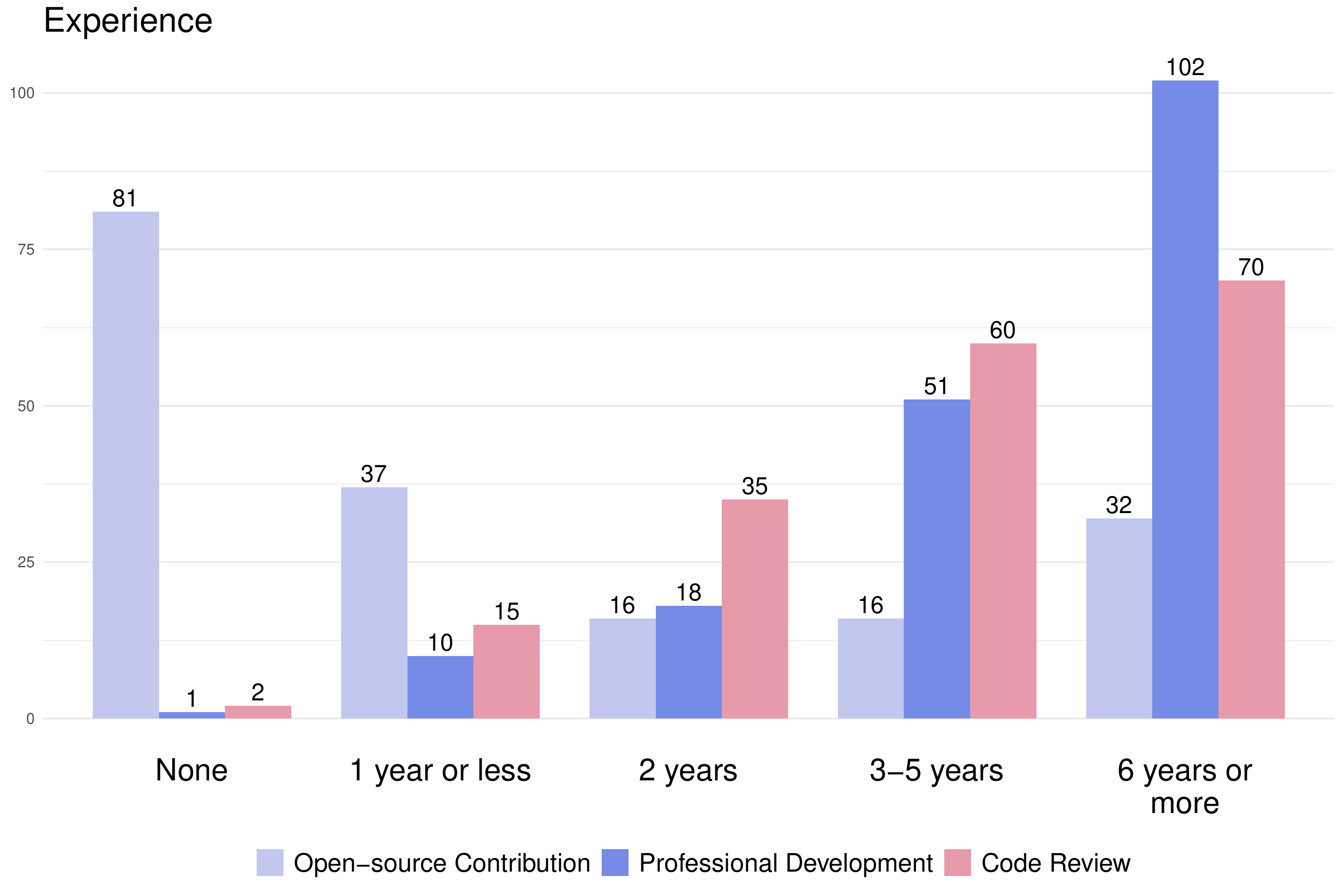}\\ 
	\vspace{0.6cm}
	\includegraphics[width=1\columnwidth]{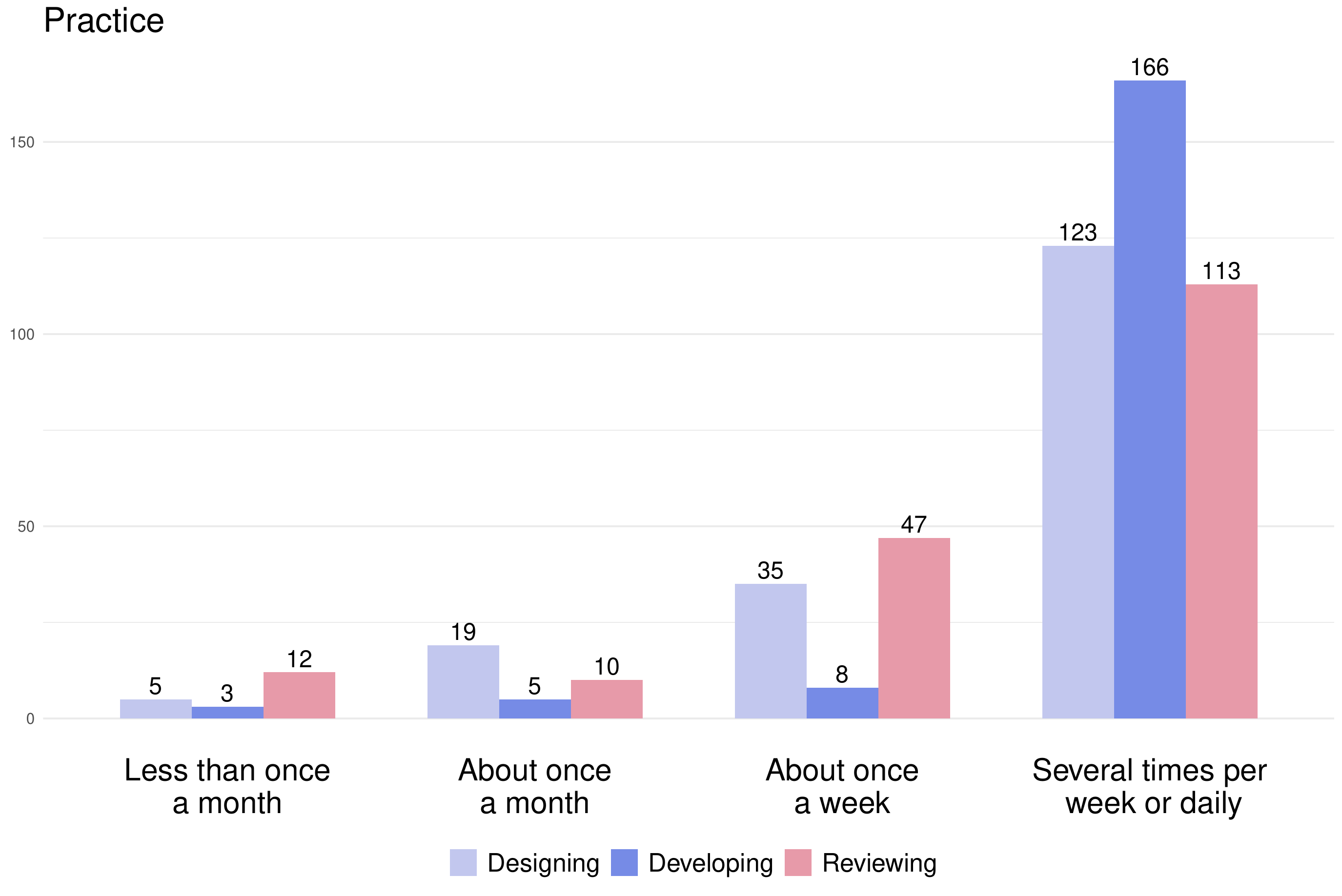} 
	\caption{respondents' demographics (absolute numbers).}
	\label{fig:demo}
\end{figure}

\Cref{fig:jobs} shows the current positions of respondents with an employment and \Cref{fig:demo} presents the respondents’ experience and practice.
In total, \numEmployed respondents reported being employed. Most respondents currently have a developer role (\perDevs), distributed as: front-end (\numDevsFront respondents), back-end (\numDevsBack), full-stack (\numDevsFull), mobile (\numDevsMobile), embedded applications or devices (\numDevsEmbbeded), and others (\numDevsOther). Moreover, most respondents (\perExperienceOverThree) report more than two years of professional development experience, and to program (\perOftenProgramming), to design (\perOftenDesign), and to review code (\perOftenCR) at least several times per week. 

\subsection{RQ$_1$. Security during Code Reviews}

Our first research question seeks to investigate the developer's perspective on software security and whether the developers in fact assess software security during code reviews.

\smallskip
\noindent\textbf{Security focus during code review. }%
When asked---as an open question--- what they focus on during code review, only \totalIssuesSecurity out of the \numParticipantsSurvey survey respondents explicitly mentioned security. Instead, other non-functional qualities of the code were mentioned more frequently (\eg \totalIssuesPerformance respondents reported performance as an issue they focus on during code review).
However, after being prompted about vulnerabilities, \totalAlwaysConsiderAgreeOrStrongly respondents reported to \emph{always} consider security during reviews. Most survey respondents (\perCanDecideAgreeOrStrongly) reported they can decide what aspects/issues to inspect during reviews (\ie in the end, ensuring security during reviews is their choice), and \perLookStrongOrAgree of them state it is their responsibility to look for vulnerabilities during reviews. 

During the interviews, two interviewees mentioned, without being prompted, to focus on security. \interviewee{4} explained: ``There are collateral effects that can be undesirable for the change\ldots like performance, security, maintainability, testability, extensibility. All these things that can be tested and generally are tested.'' 
\bluetext{Eight interviewees reported to be familiar with the term software vulnerability and provided their own definition, which was in line with ours in all cases (see \Cref{method:interviews}). \bluetext{This was also the case of \interviewee{1}, who reported to have a security background.}
The two remaining interviewees initially reported no familiarity with the term but agreed with the definition we provided. One of them further explained that they were initially unsure about the correct definition of the term, but our definition confirmed their initial opinion.}

After being prompted about software vulnerabilities, five more interviewees highlighted the importance of considering security during code reviews. Out of these five interviewees, \interviewee{7} stated: ``I consider it a lot, because I already worked for companies that we had problems related to it: the security of the data, big companies.'' 
\bluetext{Survey participants and interviewees used various terms in their answers refer to the application's security. This was the case of the previous answer given by \interviewee{7} where they mentioned ``security of the data'' when asked whether they consider security during code reviews, a follow-up question to the discussion on the software vulnerability definition. In fact, vulnerabilities in the code may lead to insecure data; for instance, a Use of a Broken or Risky Cryptographic Algorithm (CWE 327)~\cite{CWEBrokenCrypto} flaw in the code can lead to the exploit of sensitive data.} 

Moreover, two interviewees reported using automatic tools to detect software vulnerabilities, not only during code review but overall during development activities. For instance, \interviewee{2} said: ``this is automatically checked. There is a tool [that] is part of the automatic check, if that change that I am introducing, if it is exposing any credential, if there is any hardcoded password, if I let any security permission leak.'' 

\smallskip
\noindent\textbf{Participants' familiarity with security. }%
The familiarity of our study's participants with security and its concepts plays an important role in our results' interpretation. Among the survey respondents, \totalNotFamiliarSecurity stated to be unfamiliar or unsure about the term software vulnerability, while \totalFamiliarSecurity reported being familiar with it. 

Two interviewees reported to not be familiar with the term software vulnerability. Yet, they demonstrated to be security-aware after listening to the software vulnerability definition. Except for \interviewee{1} (who reported to have worked with security), the interviewees pointed up not being security experts. \interviewee{5} explained: ``[security] is a topic that most of engineers tend to not look too much into it.'' 

\smallskip
\roundedbox{
	Developers are familiar with the term software vulnerability and acknowledge the importance of ensuring security during code reviews. Yet, the vast majority of participants mentioned security as a key focus during code review only after we explicitly asked them about this topic.}

\smallskip
\noindent\textbf{Developers' assumptions about application's security. }%
The interview format allowed us to reveal potentially dangerous assumptions developers make. 
In total, seven interviewees assumed that security is the responsibility of another software component or team. For instance, \interviewee{1} explained: ``I do not worry much [about security] \ldots~I am supposing that another part of the system already dealt with it, like already dealt with the security and, at this point, I am relatively secure and should not worry much about it.'' \interviewee{9} stated: ``I am more a front-end developer, we deal less with security matters. I think [considering security during code reviews] should be more for the back end, right? They deal more with data.'' In addition, \interviewee{6} said ``I am a back-end engineer \ldots~I am not a front-end person, I do not have front-end expertise, so verifying accessibility [vulnerabilities], for example, is something I could not easily do.''

Two interviewees revealed that because they develop internal code, they do not extensively consider security during code review. For instance, \interviewee{3} said: ``I think about it for like 3 seconds and I see that there is nothing to worry about. \ldots again the code I wrote most, like 95\% of the time, is internal code that will never go outside.''  

\bluetext{In contrast to the interviewees, most survey participants reported that ensuring security is part of their jobs. In total, \totalDevRespStrongOrAgree (\perLookStrongOrAgree) survey respondents agree with the statement:``It is my responsibility to look for vulnerabilities during code reviews.'' In addition, 166 agree with the statement: ``It is part of the developer's job to ensure the security of the application.'' Yet, the difference between interviewees and survey respondents may also be explained by having prompted the latter about vulnerabilities (see \cref{sec:methog:survey})}.

\smallskip
\roundedbox{Interviewees stated to disregard security aspects during code reviews due to their assumptions about the security dynamic of the application they develop.} 

\subsection{RQ$_2$. Support to Ensure Security on Reviews} 
Our second research question seeks to investigate the support that companies and projects provide to the software security assessment during code reviews.

\begin{figure}[b] 
	\centering
	\includegraphics[width=1\linewidth,]{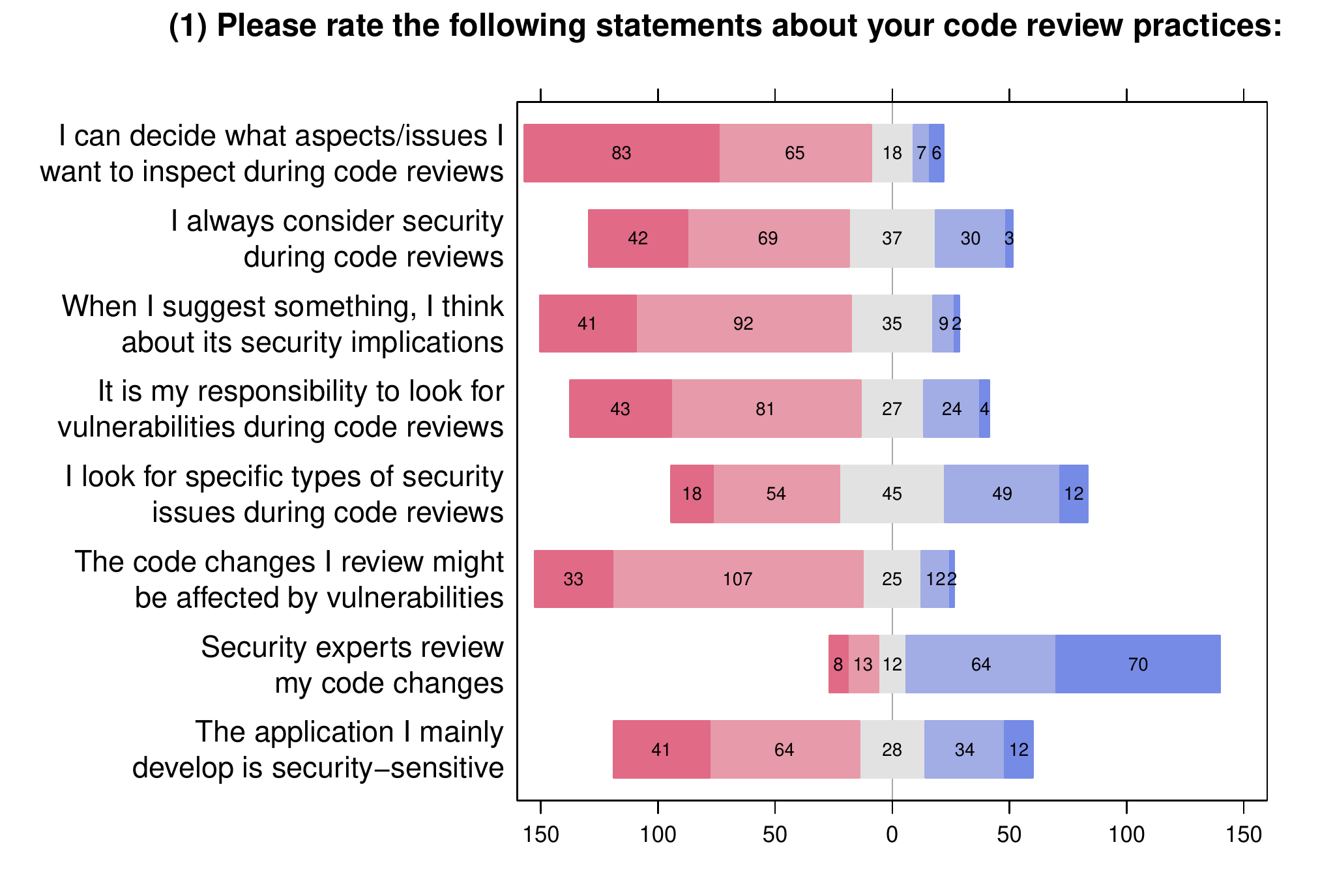} \\
	\includegraphics[width=1\linewidth,]{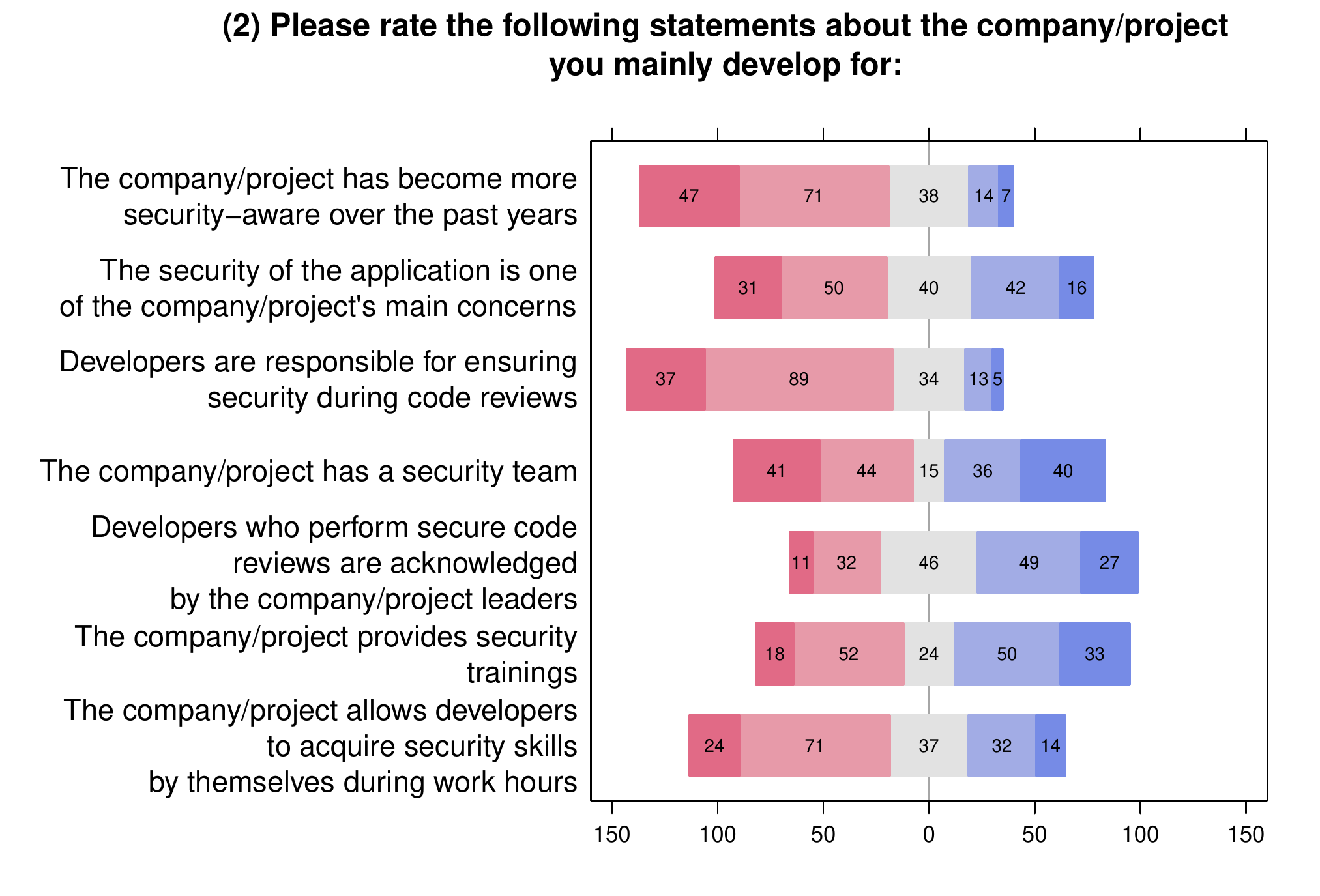}
	\includegraphics[width=1\linewidth,]{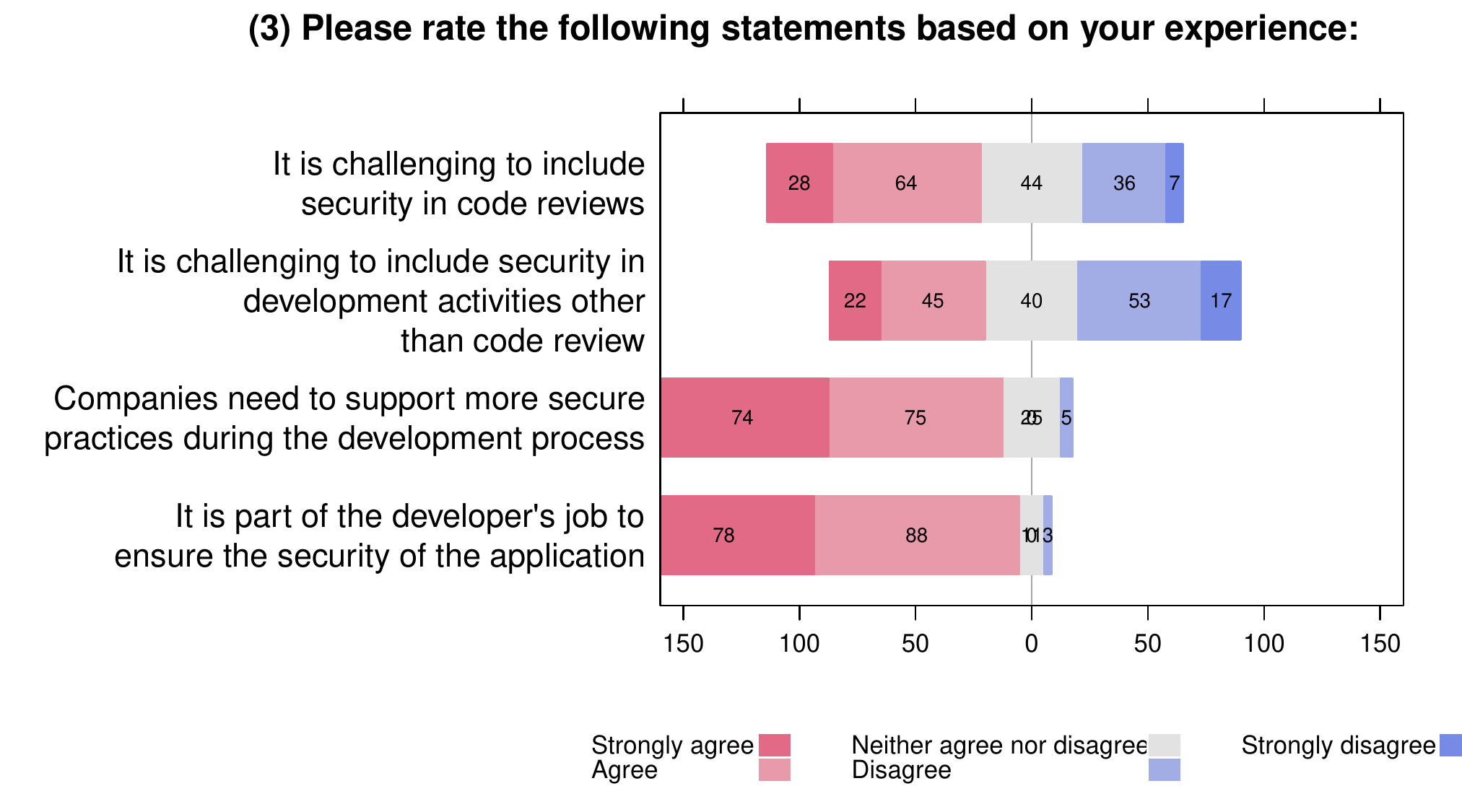}
	\caption{Participant's scaled rates about their experiences, practices and company/projects culture.}
	\label{fig:likert}
\end{figure}

\smallskip
Most survey respondents (\perCRvalue) reported to review code because they see value in code review, while only \totalCRtold of the respondents reported they conduct this activity because they are told to do so (\ie due to company/team policy). 
\Cref{fig:likert} presents the survey respondents' scaled rates about their experiences, practices, and company/projects culture. 
Overall, most survey respondents (\perSensitiveAgreeOrStrongly) reported that the main application they develop for is security-sensitive; yet, fewer (\perMainConcern) respondents reported their companies and projects to consider security as a main concern. In total, \totalHasSecurityTeam (\perHasSecurityTeam) respondents reported their company/project to have a dedicated security team, and only \perSecReviewtrongOrAgree (\totalSecReviewStrongOrAgree) of the respondents have their changes reviewed by security experts. 

\bluetext{In total, \totalAffectedStrongOrAgree (\perAffectedStrongOrAgree) respondents reported that the code changes they review might be affected by software vulnerabilities. However, only \perSecReviewtrongOrAgree of the code changes are reviewed by security experts. }
For \perDevRespOrAgree (\totalDevRespStrongOrAgree) of the respondents, developers are responsible for ensuring security during code reviews in their company/project. However, only \perACK (\totalACK) of the respondents feel like the companies and project leaders acknowledge developers that perform secure code reviews. 

Furthermore, \perProvidesSecurity (\totalProvidesSecurity) of the survey respondents reported that their company/project provides security training, but \perAllowsTraining (\totalAllowsTraining) of the companies/projects allow developers to acquire security skills by themselves during work hours. \bluetext{For \perDoMore of the respondents, companies need to support more secure practices during the development process, including during code review.}

Two interviewees commented about how their companies became more security-aware over time. For instance, \interviewee{7} told us about how their company started as a small business and how its growth impacted the security of the application: ``When I joined, the application supported very few devices \ldots~with very low sales, so it was not such a big concern, but [since we had an equipment] used by over 60,000 people \ldots~we started to have a certain kind of care. We [added] more security in the application.''

Three interviewees reported that their companies have dedicated security teams. However, security reviews of every change might not be viable: ``Unfortunately, I do not think we have enough people that are really expert in security to be able to do a scrutiny security verification in all code changes we have. \ldots It is more scalable that you give a prototype to a security team that will do an analysis to see if any attack pattern is there'' (\interviewee{6}). 

\smallskip
\roundedbox{Respondents are expected to ensure security during code reviews. However, they are not acknowledged by companies/project leaders when they do it.}

\subsection{RQ$_3$. Challenges to Ensure Security}

Our third research question seeks to understand the challenges developers face when ensuring software security during code reviews. In addition, we also explore developers' recommendations on how to mitigate these challenges.

\begin{figure}[b]
	\centering
	\includegraphics[width=1\columnwidth]{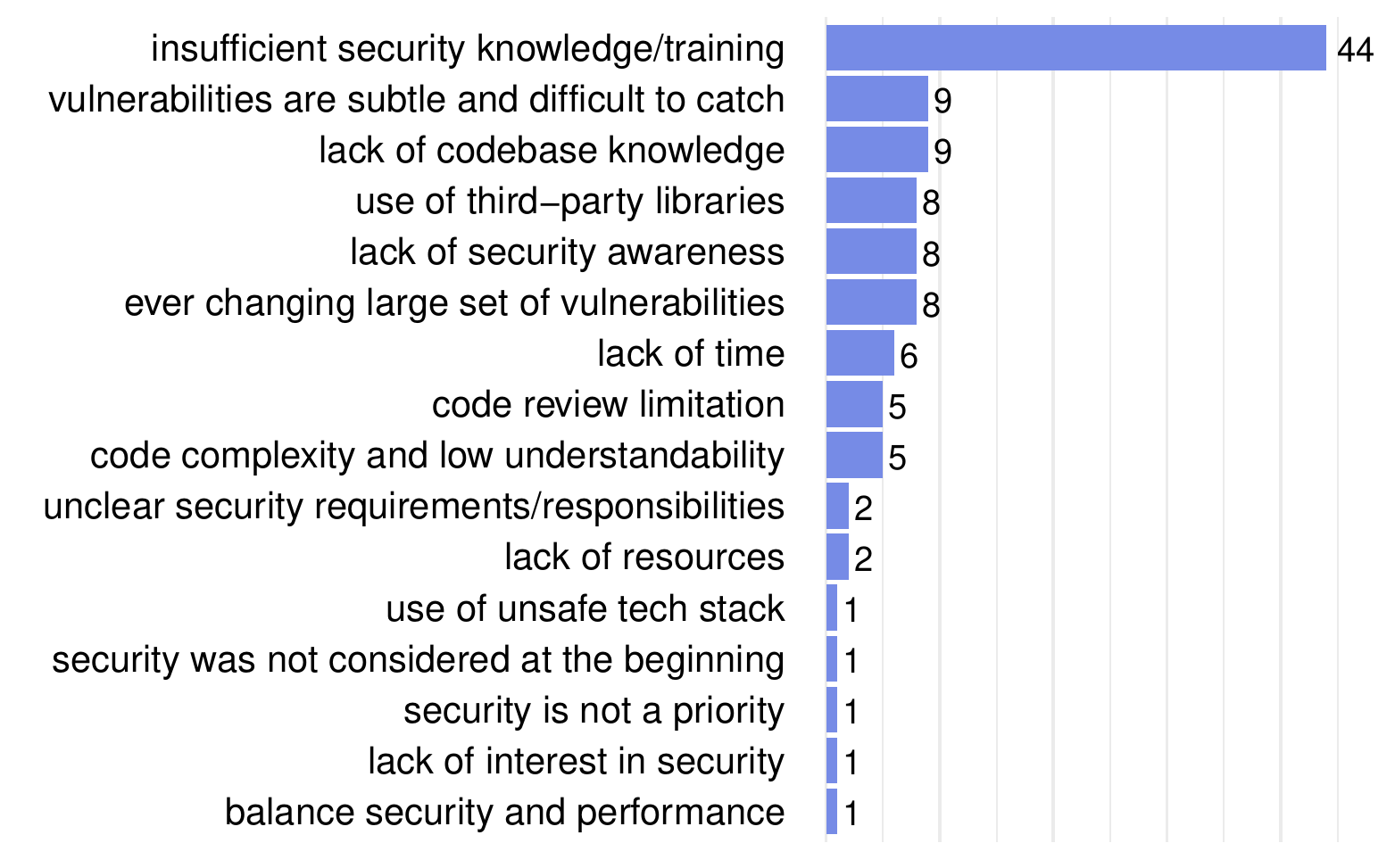}
	\caption{Challenges developers face concerning security during code reviews.}
	\label{fig:challs}
\end{figure}

When asked whether it is challenging to ensure security during code review, seven interviewees agreed. More than that, ``it can be hard even for experts to ensure security during reviews'' (\interviewee{7}). In the survey, half (\perChallCROrAgree) of the respondents reported finding it challenging to ensure security at review time, while fewer \perChallAllRespOrAgree (\totalChallAllStrongOrAgree) \bluetext{reported it as challenging} to do it in other development activities. 

Interviewees reported lack of knowledge, the large amount of possible vulnerabilities and attack patterns, time constraints, and human factors as their main challenges. \interviewee{6} said ``Because there are many, many ways of attacking, a lot more that a person who is not a specialist in the field will be able to check (...) the variability of security problems is too extensive. Because of that, it would take too much time for a nonspecialist person to verify it, besides verifying what the change needs to actually do.'' We asked survey respondents to tell us the challenges they face concerning security during code reviews as an open-text question. Respondents reported similar reasons as to the interviewees. Survey respondents mentioned lack of knowledge and time as the main reasons they find it challenging to ensure security during code reviews. For instance, the set size of possible vulnerabilities was also mentioned in the survey. \surveyR{172} explained: ``It is difficult to consider all of the possible security vulnerabilities, especially in larger ecosystems where a library may have a vulnerability that is not visible on the top layer.'' \Cref{fig:challs} shows the challenges survey respondents reported to face concerning security during code reviews. 

\smallskip
\noindent\textbf{No struggle to ensure security. }%
In total, four interviewees denied struggling to ensure software security during code reviews. For instance, \interviewee{2} said: ``I do not think [I struggle] because this is something that already existed, it is already part of the company.'' In addition, \interviewee{6} said ``I never felt like struggling. I think what already happened, that sometimes happens, is that because I am not a specialist, I end up having doubts about the security of something that I am seeing and I suggest that change goes through a review a bit more specific with a security team.''
On the other hand, three interviewees said to struggle in this activity. For example, \interviewee{7} said: ``Yes, a lot, because, although I know some points where I know the security issue is critical, I do not know everything. So, sometimes what I said before ( [small things]) go by unnoticed because sometimes I do not even know that is a point that can be exploited.'' 

Two of the interviewees also reported to struggle to include security in other activities of the development lifecycle: ``Security is something that we try to have in mind, but sometimes, I do not know, sometimes you need to deliver stuff for yesterday. Sometimes there is something in the middle of your way. Sometimes there is a certain hurry that [security] ends up unnoticed'' (\interviewee{7}). 

\smallskip
\roundedbox{Developers deny struggling to ensure security during code reviews. Moreover, doing it depends on the change's size, complexity and context.}

\smallskip
\noindent\textbf{Insufficient security knowledge and training. }%
In our survey, the most frequently mentioned challenge (mentioned by \totalChallLackKnowledge respondents) is insufficient security knowledge and training. For instance, \surveyR{35} said: ``lack of knowledge about the current vulnerabilities on the language/ framework/libraries of choice,'' and \surveyR{86} explained that ``developers often lack the low-level and/or platform expertise necessary to detect some vulnerabilities.'' \surveyR{15} explained: ``[security assessment] needs specialized knowledge (exploit techniques) and it can be hard to convince colleagues that there is a security [vulnerabilities] in their code without doing a demo if they do not have the necessary knowledge.'' Survey respondents also mentioned that vulnerabilities might not be easy to identify during the review: "Vulnerabilities are not always obvious just from looking at code, [they] sometimes slip through the cracks even when best practices are followed'' (\surveyR{86}).

Respondents were also skeptical about developers' abilities to learn and perform secure code reviews. For instance, \surveyR{154} explained ``Dedicated security experts are not economically viable for most smaller companies / teams / project. Some training might help but it is probably an unrealistic for all developers to become competent enough in IT security.
Addressing this issue requires better tool support (static code analysis) and improved frameworks / libraries / API, so that developers have less `opportunity' to mess up.'' 

\medskip
\noindent\textbf{Vulnerabilities are hard to detect and keep up with. }%
The nature of software vulnerabilities also impacts its detection. In total, \totalChallSubtle survey participants reported that vulnerabilities are subtle and difficult to detect, even when the best practices are followed. For instance, \surveyR{86} explained that ``vulnerabilities are not always obvious just from looking at code,'' and \surveyR{154} said that ``some vulnerabilities are very subtle (yet dangerous) and therefore easy to overlook. The code actually does not look `bad'.'' In addition to that, the large amount of vulnerabilities and the constant discovery of new ones turns keeping up with security a hard task. For example, \surveyR{31} reported the ``ever-changing development landscape, new vulnerabilities are found in third-party packages every day'' as a challenging factor to ensure security during review.

\medskip
\noindent\textbf{Lack of codebase knowledge. } %
Lack of codebase knowledge was mentioned \totalChallCodebase times. Survey respondents reported lack of knowledge about applications' architecture, domain, and components interaction as factors that influence the detection of vulnerabilities during reviews. For instance, \surveyR{39} explained ``I am not a security expert; my reviews will never be airtight on a security level. The level of threat is hard to determine since company-wide network policies and other teams' applications are also part of the ecosystem with details that are not necessarily known.'' In addition, \surveyR{42} said ``lack of knowledge of the overall domain (I work in a heavily regulated financial environment and I am relatively junior) with many existing security measures. Awareness of the interactions between these systems is desirable when inspecting security.''

\smallskip
\noindent\textbf{Use of third-party libraries. }%
Lack of knowledge also applies to third-party libraries used in the application. The use of third-party libraries was mentioned by \totalChallThird survey respondents as a challenge they face when ensuring security during code reviews. \surveyR{86} said ``Vulnerabilities are not always present in the code we maintain, often they are library/platform/integration issues." 

\medskip
\noindent\textbf{Security awareness at review time. }%
The results of \textbf{RQ$_1$} show that developers are familiar with security but do not focus on it during code reviews. Some developers acknowledge this scenario. In total, \totalChallAwareness survey respondents mentioned the lack of security awareness during code reviews as a challenging factor to detect vulnerabilities during the review. For instance, \surveyR{16} reported ``general lack of security awareness from most developers'' as a challenge they face. Moreover, \surveyR{147} said that ``[security] is typically not on the forefront of the mind of the author.''

\medskip
\noindent\textbf{Time and security interest. }%
Lack of time (\totalChallTime respondents), interest in security (\totalChallInterest), and resources (\totalChallResources) were also mentioned as challenge factors developers face to ensure security at code review time. \surveyR{62} said ``[considering security during code reviews] is too time-consuming.'' In addition, \surveyR{132} stated: ``Lack of time/interest/ funding for security-centric changes leads to security being an afterthought.'' Finally, \surveyR{80} wrote: ``So many changes happen that it can be hard to see how a small change affects the security. Security has not been thought of at the start and was hacked in.'' 

\medskip
\noindent\textbf{Human factors. }%
Survey participants also mentioned human factors, such as fatigue and focus, as factors that impact the detection of security issues during code reviews: ``It is time pressure, fatigue, focus, and all these things, because if you have to review lots of codes, you will end up lowering your attention. If you are constantly interrupted during a code review, when you need to assemble the system in your head, assemble the change in your head along with the system and seeing how it works, that can get in the way'' (\interviewee{4}).

\medskip
\noindent\textbf{Code review limitations. }%
In total, \totalChallCRlimitations survey respondents also mentioned some limitations that code review poses to ensuring security. For instance, \surveyR{64} mentioned ``too much code to review'' as a challenging factor to ensure security during reviews, and \surveyR{112} said: `` [I am] not always sure what to look for, nor sure if it is my responsibility instead of our security team’s responsibility. Code reviews also show only a limited view of an application, and proper understanding of security flaws may require seeing how code interacts across the whole application.'' \surveyR{86} wrote: ``spotting vulnerabilities often requires deep understanding of the subject code which is not really required of code reviewers.'' 

\medskip
\roundedbox{Insufficient security knowledge is the mainly reported challenge in ensuring security during code reviews. Knowledge regarding the application and its components, third-party libraries, human factors, and code review limitations also hinder security assessment at review time.}

\section{Limitations}

Assessing the validity of explorative qualitative research is challenging~\cite{onwuegbuzie:2007,golafshani:2003}. Despite our best efforts, some limitations exist; in the following, we explain how we tried to minimize them.

To mitigate the possible limitation from missing control over subjects, we included questions to characterize our sample, such as experience and role (Block 4 in \Cref{sec:methog:survey}). 
We also removed participants who did not complete the experiment and reported not performing code reviews. We manually analyzed participants that incorrectly answered the qualification questions.

During the interviews, we might have led interviewees to provide more desirable answers~\cite{hildum:1956}. To mitigate this issue, we challenged and triangulated our findings with the survey results. 
Our study participants might have given socially acceptable answers to appear in a positive light. To mitigate this social desirability bias~\cite{furnham:1986}, we informed participants that the responses would be anonymous and evaluated in a statistical form. 
To mitigate question-order effect~\cite{sigelaman:1981} that might have led the survey respondents to specific answers, we randomized the order of answers of the multiple-choice questions and the statements of the scale rate questions.

We collected qualitative data from the interviews and the survey to understand developers’ main challenges to ensure security during code reviews. We used the data to get insights into their perceived challenges and what they would do/recommend to do to mitigate such issues. We used different measurement techniques to mitigate \emph{mono-method bias}~\cite{Cook:1979}: we obtained qualitative results by employing card sorting on participants’ responses to the survey's open-text questions (Block 3 in \Cref{sec:methog:survey}). We also used this technique on the interviews' transcripts. 

We only collected data through interviews and surveys, which may not provide the full picture of developers' perceptions. To mitigate this limitation, one can examine code, design documents, issue tracking system contents, and other repositories. 
We hope that future research can be inspired by our results and triangulate selected findings with other data sources.

To obtain a diverse sample of participants, we invited software developers from several countries, organizations, education levels, experiences, and backgrounds. Even though our sample comprises several types of software developers and we found a large agreement concerning their perceptions, we cannot claim it is representative of all developers. If the study is repeated using different participants, the results may be different. 

Participants could freely decide whether to participate in the study or not (self-selection). They were informed about the survey’s topic (\emph{code review}), an estimated duration for the participation, and offered a donation to a charity institution to encourage their participation. This could have biased the selection of participants as only participants who could spare enough time or were interested in the incentive might have participated. We tried to mitigate this risk by advertising through various channels.

\section{Discussion}

We discuss how the findings on the developer's perception of software security code reviews can be used to better support the secure development of applications, and we outline opportunities for future work.

\medskip
\noindent\textbf{A change in developer's perspective on security.}
In the past, developers have shown an ``it is not my responsibility'' attitude towards software security~\cite{Xie:2011} and limited adoption of Secure Development of Applications (\sda)~\cite{woon:2007}. 
However, software security seems to be gaining more space among developers’ responsibilities. In fact, a study~\cite{Christakis:2016} reported that developers care more about security than other reliability issues. 
Following this trend, our results suggest a change in developer's attitude---they state it is their responsibility to ensure security during code reviews. 

\medskip
\noindent\textbf{Security needs to be better motivated.}
Even though developers seem to have changed their attitude towards security, recent research~\cite{braz:2021,braz:2022} has also provided initial evidence that security is not in the developers' mind ``by default'' when reviewing code. Our results corroborate this evidence: We found that developers do not have security in mind when describing their code review practices. A reason for this might be that developers lack motivation to be concerned about security during reviews. Indeed, companies and projects expect developers to ensure the security of their applications and allow them to acquire security knowledge during work hours, but they do not provide the means for that, \eg security training and workshops. Moreover, leaders do not acknowledge developers that do secure code reviews.

This situation raises questions on the effectiveness of how companies motivate developers in the software development process, especially at code review time.
Organizations may consider incorporating explicit reward systems for developers who ensure security in the applications. For instance, the Vulnerability Reward Program~\cite{GoogleVRP} rewards any bug bounter who report vulnerabilities in Google-owned and Alphabet subsidiary web properties. Inspired by this initiative, companies and projects may create programs to reward developers who detect vulnerabilities through code review. 

\medskip
\noindent\textbf{Assumptions may hurt security. }%
Developers make security assumptions during code reviews. For example, \citet{braz:2022} reported that reviewers justified not detecting vulnerabilities because they assumed the change author had already considered the application's security. In line with their findings, our study participants assume that security is the responsibility of another application's component or team. For instance, back-end developers reported assuming security as a front-end responsibility, while front end developers reported the opposite. However, developers need to be careful and spread awareness around these assumptions so that they do not end up hurting the application's security. 

Some developers also thought their code is not security-sensitive because the software is intended for internal use. However, malicious and inadvertent activities may happen inside an organization. For example, acknowledging these scenarios, Microsoft has an insider risk management compliance solution that helps minimize internal risks by enabling clients to detect, investigate, and act on this type of activities~\cite{microsoft-risks}.
Developers should be aware that internal code may still be vulnerable to malicious attackers that impersonate employees.
Organizations may consider raising awareness on this issue and incorporating more strict software security policies into their development process to create a different attitude. 

\medskip
\noindent\textbf{Security can be learned. }%
Participants mainly reported insufficient security knowledge as a challenge when considering security during code review. 
Our results align with previous work \cite{braz:2021, braz:2022}, in which developers reported lack of knowledge as the reason for not finding vulnerabilities in a code review task. 
These results strengthen the questions on the security education (or lack thereof) that developers are receiving. 
To create a different approach, educational institutions may introduce or give more attention to security in the software engineering undergraduate and graduate courses. For instance, in 2004, a team from the US Naval Postgra\-duate School won the ``Capture the Flag'' tournament at Defcon, the world's most popular hacker convention~\cite{defcon}. 
To learn security, students and developers must be able to switch from their traditional conditioning to the attacker’s way of thinking~\cite{Bratus:2007}. Studies can be carried out to determine how to better educate students on software security.

Furthermore, to overcome developers' insufficient security knowledge, companies and projects may provide security training and allow time for learning. The free resources provided by OWASP~\cite{owasp} can be used to educate developers on security. For instance, organizations can adopt regular talks about the OWASP Top 10 List~\cite{owaspTop10} to keep developers aware of the riskiest vulnerabilities, as well as training and seminar about new emerging vulnerabilities. 

In addition, software developers can use Massively Open Online Courses (MOOCs) to learn security beyond the setting of their company/project. MOOCs are online courses that, additionally to traditional course materials, provide interactive courses with user forums or social media discussions to support community interactions among students, professors, and teaching assistants, as well as immediate feedback to quick quizzes and assignments~\cite{moocs-wiki}. Although previous research~\cite{theisen:2016} has provided evidence that on-campus security courses still have better results than security MOOCs, developers can still benefit from them. For instance, our results show that companies allow developers to learn security during their work hours. These developers may follow MOOC activities from the office. Further studies can be conducted to investigate how to improve MOOCs to better support developers learning software security. 

\medskip
\noindent\textbf{A little help from the experts.} %
Most of our participants reported that they develop security-sensitive applications, but their code is rarely reviewed by security experts. This is a missed opportunity because code review can be used as a learning tool~\cite{Bacchelli:2013}. Including an expert in the review may not only increase the security of that specific change, but developers can make use of the knowledge sharing during the review to learn, \eg it can be a way to do onboarding in the security field.

The 2014 Cisco Annual Security Report estimated a potential shortfall of a million security professionals globally~\cite{Cisco}. \bluetext{In 2022, the report of the World Economic Forum~\cite{WEF-report} stated: ``There is an undersupply of cyber professionals --- a gap of more than 3 million worldwide --- who can provide cyber leadership, test and secure systems, and train people in digital hygiene.'' }
This way, even large companies do not have enough security experts to review every change made by developers. 
In fact, only one-third of our survey respondents reported having a security team in their companies/projects.
 As a solution, teams can select a developer to increase their security expertise and propagate their knowledge to other team members through code reviews. 
OWASP~\cite{owasp} suggests that companies not only have a security team but also a further group of developers with an interest in security that can act as team local security Subject Matter Experts (SMEs) taking part in the code reviews~\cite{owaspChecklist}.
Further studies can be designed and conducted to evaluate the effect of learning security in reviews. 

\medskip
\noindent\textbf{The right change to the right reviewer. } %
When not enough experts are available to review all code changes, the most security-critical changes can be selected and sent to the experts for a more scrutiny security analysis. Tools and processes can be developed to recognize security-critical changes and automatically assign security experts as reviewers. Previous research~\cite{Thongtanunam:2015, Rahman:2016,Balachandran:2013,kovalenko2018does} has investigated reviewer recommendation approaches, including recommendations based on cross-project work experience of potential reviewers and estimation of their expertise in specific technologies~\cite{Rahman:2016}. These approaches may be adapted to take security aspects into consideration. For example, they may incorporate security static analyzers, such as SonarQube~\cite{SonarQube} and Snyk Code~\cite{Snyk}, to measure the change's security-critical level.

\medskip
\noindent\textbf{Code review is not perfect. }%
Code review is considered a valuable tool for finding defects and improving software quality~\cite{Ackerman:1989}. In its current form~\cite{Bacchelli:2013}, it requires short-term reactions from reviewers on the code changes~\cite{sadowski2018modern}. In our studies, participants reported limitations specific to the code review process that hinder security assessment of code changes. For instance, developers reported that ensuring security is a time-consuming activity and do not have enough time for it due to their high workload. In addition, the security impact of code changes can spread in the application and not be limited only to the diff shown in the review. This way, to identify possible software vulnerabilities, developers need to have a good understanding of the application's components interaction, architecture design, and third-libraries usage, which does not always happen, thus often leaving software vulnerabilities undetected. 
Previous studies have reported the benefits of visualization techniques for developers~\cite{khaloo:2017,eick:1992,mattila:2016}, such as supporting them in better understanding or navigating the code. Future research can investigate whether developers can improve their understanding of the security impact of the code changes through the use help of code change visualization tools, such as \textsc{Code Park}~\cite{khaloo:2017} and \textsc{CHANGEVIZ}~\cite{Gasparini:2021}.

\section{Conclusions}

In this paper, we investigated the developer's perspective on assessing software security during contemporary code reviews, the support companies and projects provide to this process, and the challenges they face in this task.
To this aim, we interviewed ten professional developers and surveyed \numParticipantsSurvey practitioners. 

Our results show a change in the developer's attitude towards security -- they acknowledge it is their responsibility to ensure security during code reviews. Yet, we also find that developers do not have security in mind when describing their code review practices, thus suggesting that security is not one of their main priorities when reviewing code. Moreover, developers make assumptions about the dynamic of the application for which they are reviewing code. These assumptions may lead them to disregard security aspects during their code reviews.

Our findings indicate that companies may need to change their approach towards security: They expect developers to ensure the security of their applications but provide little support and incentives to do so. 
Developers deny struggling to ensure security during code reviews. Yet, they report insufficient security knowledge, lack of knowledge regarding the application and its components, third-party libraries, human factors, and code review limitations as challenging factors that hinder security assessment at review time. 

Our findings raise questions on the effectiveness of current methods employed in the software development process to ensure security, especially at code review time. To achieve secure and maintain secure applications, developers need to be educated about software security, and organizations need to improve how they motivate developers about it. Moreover, strategies to better support developers when ensuring security during code reviews should be further investigated to establish and improve the code review process.

\begin{acks}
The authors would like to thank the anonymous reviewers for their thoughtful and important comments, which helped improving our paper. The authors gratefully acknowledge the support of the Swiss National Science Foundation through the SNSF Projects No. 200021M\_205146 and PZ00P2\_186090.
\end{acks}

\bibliographystyle{latex/ACM-Reference-Format}
\bibliography{fse22_security-main}

\end{document}